\RequirePackage{fix-cm}
\documentclass[twocolumn,epjc3]{svjour3}
\RequirePackage{graphicx,subfigure}
\usepackage[shortcuts]{extdash}

\begin{document}
\title{A multi-isotope $0\nu2\beta$ bolometric experiment}
\titlerunning{multi-isotope bolometric experiment}

\author{
A.~Giuliani\thanksref{addr1,addr2,e1}\and
F.A.~Danevich\thanksref{addr1,addr3}\and
V.I.~Tretyak\thanksref{addr3}
}
\thankstext{e1}{e-mail: andrea.giuliani@csnsm.in2p3.fr}
\institute{CSNSM, Univ. Paris-Sud, CNRS/IN2P3, Universit\'e Paris-Saclay, 91405 Orsay, France \label{addr1}
\and
DISAT, Universit\`a dell'Insubria, 22100 Como, Italy \label{addr2}
\and
Institute for Nuclear Research, 03028 Kyiv, Ukraine \label{addr3}
}


\maketitle

\begin{abstract}

There are valuable arguments to perform neutrinoless double beta ($0\nu2\beta$) decay experiments with several nuclei: the uncertainty of nuclear-matrix-ele\-ment calculations; the possibility to test these calculations by using the ratio of the measured lifetimes; the unpredictability of possible breakthroughs in the detection technique; the difficulty to foresee background in $0\nu2\beta$ decay searches; the limited amount of isotopically enriched materials. We propose therefore approaches to estimate the Majorana neutrino mass by combining experimental data collected with different $0\nu2\beta$ decay candidates. In particular, we apply our methods to a next-generation experiment based on scintillating and Cherenkov-radiation bolometers. Current results indicate that this technology can effectively study up to four different isotopes simultaneously ($^{82}$Se, $^{100}$Mo, $^{116}$Cd and $^{130}$Te), embedded in detectors which share the same concepts and environment. We show that the combined information on the Majorana neutrino mass extracted from a multi-candidate bolometric experiment is competitive with that achievable with a single isotope, once that the cryogenic experimental volume is fixed. The remarkable conceptual and technical advantages of a multi-isotope investigation are discussed. This approach can be naturally applied to the proposed CUPID project, follow-up of the CUORE experiment that is currently taking data in the Gran Sasso underground laboratory. 
\end{abstract}



\section{Introduction}

Neutrinoless double beta ($0\nu2\beta$) decay is considered \linebreak[4] as one
of the most promising ways to investigate the properties of neutrino
and weak interactions, to test lepton number conservation, and in general to probe the Standard Model (SM) of particle physics
\cite{Barea:2012,Deppisch:2012,Rodejohann:2012,Bilenky:2015,Pas:2015,DellOro:2016,Vergados:2016}.
Despite the almost seventy-year-long history of $0\nu2\beta$ decay
search, this elusive process is still not observed: the
most sensitive experiments provide limits on the effective
Majorana neutrino mass of the electron neutrino on the level of $\lim
\langle m_{\nu}\rangle \sim 0.1-1$ eV (see the status of the
$0\nu2\beta$ decay search in the reviews
\cite{Deppisch:2012,Rodejohann:2012,Bilenky:2015,Pas:2015,DellOro:2016,Vergados:2016,Elliott:2012,Giuliani:2012a,Cremonesi:2014,Gomes:2015,Sarazin:2015}
and the recent experimental results
\cite{EXO-200,CUORE-0,CUORE,NEMO-3,GERDA,KamLAND-Zen}).

The exploration of the inverted neutrino mass hierarchy ($\langle
m_{\nu}\rangle \sim 0.02-0.05$ eV) is a challenge of the
experiments in preparation or R\&D stage. Such a sensitivity can
be reached in searches able to detect an extremely low activity
with half-life in the range $T_{1/2} \sim 10^{26}-10^{27}$ yr for
the most promising candidates. However, taking into account
the possible quenching of the axial vector coupling constant $g_A$
\cite{Barea:2013,DellOro:2014,Suhonen1:2017,Suhonen2:2017}, the experimental program should
foresee the possibility to go towards an-order-of-magnitude higher
half-life sensitivities, of the order of $T_{1/2} \sim
10^{27}-10^{28}$ yr. The accomplishment of such an ambitious plan
requires the construction of detectors containing a large number of
$2\beta$ active nuclei ($10^{27}-10^{28}$ nuclei, $\sim 10^3-10^4$
moles of isotope of interest), with an as-high-as-possible detection
efficiency and an extremely low (ideally zero) radioactive
background, and able to distinguish the effect searched for. The last
requirement implies, in particular, an as-high-as-possible energy
resolution --- to identify the effect unambiguously and to avoid
the presence of a tail from the SM-allowed two-neutrino double beta ($2\nu2\beta$)
decay in the region of the $0\nu2\beta$ peak (region of interest, ROI) --- and time resolution --- to
avoid random coincidences of events from $2\nu2\beta$ decays and other components of
the background \cite{Chernyak:2012}.

The technique of scintillating and Cherenkov-radia\-tion low-temperature bolometers
\cite{Alessandrello:1998,Pirro:2006,Giuliani:2012b,Artusa:2014,Tabarelli:2010,Poda:2017} is
one of the most suitable detection method for the next generation
$2\beta$ experiments, thanks to the high detection efficiency
($70\%-90\%$, depending on the crystal composition and
volume) and the excellent energy resolution (a few keV). In
addition, this technique suppresses the background caused by $\alpha$ radioactive contamination to a very low level thanks to the excellent particle discrimination. Following this approach, the CUPID group of interest  \cite{CUPID-RD} proposes to perform a tonne-scale bolometric $0\nu2\beta$ decay search with a sensitivity high enough to probe the inverted hierarchy of the neutrino mass and
even go towards the normal hierarchy region. This experiment would be a follow-up to the CUORE experiment with particle identification for background rejection.

There are both theoretical and experimental arguments to develop
$0\nu2\beta$ decay experiments with several nuclei. First of all,
the ambiguity of the theoretical calculations of nuclear matrix
elements (NME) of the $0\nu2\beta$ decay remains significant, not to
say that there are no experimental tests of the calculations so far.
Therefore, no one can safely suggest the nucleus with the highest decay
probability. Furthermore, data on $0\nu2\beta$ decay rate in a few
nuclei will be definitely useful to adjust theoretical
calculations of the NME since the ratio of the calculated
lifetimes is very sensitive to different models
\cite{Bilenky:2002}. From the experimental point of view, taking into account the
extremely low decay probability and the possibility that spurious effects mimick the signal searched for, an observation of $0\nu2\beta$ decay in different nuclei will be requested in case of positive evidence in one of the candidates. Indeed, there have already been many false claims of double beta decay in the history of the experimental search for this process \cite{Tretyak:2011}. 

Therefore, there is a strong call for experiments with several nuclei candidates. At the same time one can extract  competitive estimations of the Majorana neutrino mass by combining data of experiments with several different isotopes. In this paper we demonstrate that the sensitivity to the Majorana neutrino mass that can be achieved in a cryogenic experiment with several nuclei is similar to the sensitivity of an experiment utilizing only one of the candidates and exploiting the same experimental volume. Of course, the same argument can be applied to other physics parameters related to different mechanisms inducing $0\nu2\beta$ decay, although deepening this aspect goes beyond the scope of this paper. 

\section{Experimental concepts}

On the basis of the above considerations, we propose a next generation $0\nu2\beta$ decay bolometric experiment studying several isotopes simultaneously. To complete the general discussion of the previous Section, we stress that there are also practical, but not less important, reasons to implement such cryogenic $0\nu2\beta$ experiment: the production of a large amount of a given isotope (at the hundreds of kg scale) looks rather questionable, while the production, e.g., of one quarter of the same quantity for four isotopes looks more realistic, given the existing enrichment techniques and capabilities. We remark that different sets of centrifugues are used for different nuclei in the centrifugation enrichment technique (presently the only one that provides a high enough throughput). Enrichment in different isotopes can therefore be performed in parallel. A similar argument applies to crystal production, which is mandatory for the bolometric approach. Crystals containing different candidates can be grown in parallel, since each type of crystal requires specific technologies and is typically produced in a specialized laboratory or company. A multi-isotope experiment can therefore provide a remarkable reduction of the time required to build the detector, especially in the case, discussed below, in which the single module structure is the same for each type of crystal.

We will consider the candidates $^{82}$Se, $^{100}$Mo, $^{116}$Cd and $^{130}$Te, which can be studied with the scintillating (the first three ones) and Cherenkov-radiation (the last one) bolometric technique with excellent prospects. Although our considerations apply to a generic next-generation cryogenic experiment, the most natural implementation of this multi-isotope approach is in the framework of the CUPID proposal. Therefore, in the following we will refer to the four-isotope option as CUPID\=/4, while we will designate CUPID\=/1 the conventional one-isotope version preliminarily discussed in Ref. \cite{CUPID-RD}. 

As for the detector technology, we will stick to the basic configuration of the single module of CUORE \cite{CUORE-0,CUORE}, consisting of a large crystal (125 cm$^3$ in case of CUORE) coupled to a neutron transmutation doped (NTD) Ge thermistor, working as a temperature sensor, and to a Si-doped heater for detector response stabilization. Each crystal is hold by small PTFE elements inside a copper frame. The read-out electronics operates at room temperature and is based on the well-known Si-JFET technology. CUORE is taking data successfully  \cite{CUORE} and there is no reason to change substantially this experimental configuration. 

The bolometric light detector, which is not present in CUORE but will be necessary to reject the $\alpha$ background in future bolometric searches, will consist of a thin Ge disk (50 mm diameter, 0.15 mm thickness) with an NTD Ge thermistor as well. It will be operated in the Neganov-Luke mode in order to achieve a high enough sensitivity to reject $\alpha$ background in the TeO$_2$ crystals --- used in the $^{130}$Te section --- by exploiting Cherenkov radiation \cite{Tabarelli:2010,NL1,NL2}  and the two-neutrino double beta decay random coincidences in the $^{100}$Mo section \cite{Chernyak:2017}. The same Si-JFET-based read-out electronics adopted for the main crystal, and used now in CUORE, can be employed. The same type of light detectors can be coupled to all the four crystal types, but the Neganov-Luke option is not strictly necessary in the $^{82}$Se and $^{116}$Cd cases. Of course, other technologies for the light detector may be viable: a vibrant R\&D activity is ongoing on this subject \cite{Poda:2017,CUPID-RD}. 
 
We will assume an array of 1300 scintillating and Cherenkov-radiation bolometers (compatible with the experimental volume of the current CUORE cryostat or of a similar dedicated infrastructure) based on cylindrical ultra-radiopure crystals --- 5 cm diameter and 5 cm height --- enriched to 100\% in the isotope of interest. In case of the CUPID\=/1 option, the crystals will consist of only one of the following compounds: zinc selenide (Zn$^{82}$Se), lithium molybdate (Li$_2$$^{100}$MoO$_4$), cadmium tungstate ($^{116}$CdWO$_4$), or tellurium oxide ($^{130}$TeO$_2$). As for the CUPID\=/4 option, we will assume that the 1300 elements will be shared among crystals of each type. We stress that a remarkable advantage of the bolometric technique for a multi-isotope experiment is that detectors are based exactly on the same technology, are operated in the same set-up and are exposed to the same background sources.

All the four aforementioned compounds have been successfully tested as scintillating or Cherenkov-radiation bolometers, using crystals grown from enriched materials. It was shown as well that the Neganov-Luke light detectors have the requested performance \cite{NL2}. The energy resolution of the detectors at the energies $Q_{2\beta}$ were estimated from existing experimental data concerning ZnSe \cite{Beeman:2013}, Li$_2$MoO$_4$ \cite{Armengaud:2017}, CdWO$_4$ \cite{Arnaboldi:2010}, and TeO$_2$ \cite{Artusa:2015}. The detection efficiencies for the $0\nu2\beta$ decay effect were taken from Ref. \cite{Artusa:2014} (ZnSe and TeO$_2$), while for Li$_2$MoO$_4$ and CdWO$_4$ detectors the efficiencies were estimated by using the GEANT4 simulation package.

\section{Approaches to estimate neutrino mass from several $0\nu2\beta$ experiments}
 \label{sec:2}
 
A value (limit) on the effective Majorana neutrino mass can be derived from the
experimental data collected by an experiment following the CUPID\=/4 scheme. Attempts to estimate a limit on the effective Majorana neutrino mass combining data of the most
sensitive experiments are already reported in the literature \cite{Guzowski:2015,Klimenko:2017}. We propose here other approaches driven also by the quite similar performance of the cryogenic detectors foreseen in CUPID.

\subsection{Sum of counts in the region of interest}
 \label{sec:21}

Let us consider the mass mechanism for light neutrino exchange, which leads to the following expression for the half-life $T_{1/2}^{0\nu}$ of the nuclei candidates to the $0\nu2\beta$ decay:

\begin{equation}
[T_{1/2}^{0\nu}]^{-1}= G^{0\nu} |M^{0\nu}|^2 g_{A}^4 \frac{\langle
m_{\nu} \rangle^2}{m_{e}^2}, \label{eq:1}
\end{equation}

\noindent where $G^{0\nu}$ is the phase-space factor, $M^{0\nu}$
is the NME, $g_{A}$ is the axial vector coupling constant,
$\langle m_{\nu} \rangle$ is the Majorana neutrino mass, and
$m_{e}$ is the mass of electron. The number of detected $0\nu2\beta$
events $S_i$ in an $i$-th $0\nu2\beta$ experiment with $N_i$
nuclei, with a detection efficiency $\varepsilon_i$, over a live time
$t_i$ is inversely proportional to the nucleus half-life:

\begin{equation}
S_i=\frac{\ln2 N_i\varepsilon_i t_i}{(T_{1/2}^{0\nu})_i}.
 \label{eq:2}
\end{equation}

\noindent The sum of $0\nu2\beta$ events acquired by several detectors containing different $2\beta$ isotopes can be expressed as following:

\begin{equation}
 \sum S_i=\frac{\langle m_{\nu} \rangle^2}{m_{e}^2}g_{A}^4\ln2\sum N_i\varepsilon_i t_i
 G_i^{0\nu}|M_i^{0\nu}|^2.
 \label{eq:3}
\end{equation}

\noindent Therefore, the Majorana neutrino mass can be derived from several experiments by using the following formula:

\begin{equation}
\langle m_{\nu} \rangle = \frac{m_{e}}{\sqrt{\ln 2}~g_{A}^2}
\sqrt{\frac{{\Sigma S_i}}{\sum N_i\varepsilon_i t_i
G_i^{0\nu}|M_i^{0\nu}|^2}}.
 \label{eq:4}
\end{equation}


\noindent $S$ ($\lim S$) can be obtained from the analysis of the experimental data accumulated with several detectors as following (assuming a large enough number of counts):

\begin{equation}
S = \Sigma S_i \pm \sqrt{\Sigma \sigma^2(S_i)}.
 \label{eq:5}
\end{equation}


However, the method of sum of counts has a drawback in the case of different detectors performance: if all the $S_i$ values are close to 0, the biggest contribution will be given by the measurement with the highest background. Thus, the neutrino mass limit could be defined by the worst experimental case, which is not logical.

Nevertheless, as it will be shown in Sections \ref{sec:33} and \ref{sec:34}, the method is quite effective in case of experiments with similar mass of the isotope of interest,
background, efficiency and energy resolution, which is expected to be
the case of CUPID\=/4.

In Section \ref{sec:34} we will consider ``zero background" CUPID-like
experiments. Approximation of data with mainly bins with zero
counts is not robust enough. Thus, one can calculate the sum of counts
in selected energy intervals to estimate experimental sensitivities
to the Majorana neutrino mass by using only the given background and
no true signal.


\subsection{Weighted averages and errors of neutrino mass square}
 \label{sec:22}

The following method allows us to suppress the impact of measurements with high backgrounds. A value (limit) of neutrino mass can be derived from several experiments with different nuclei by using weighted averages and errors of square of the neutrino mass obtained
in different experiments as recommended by the Particle Data Group
\cite{PDG:2016} (assuming once again a large enough number of counts):

\begin{equation}
\langle m_{\nu} \rangle^2 \pm \sigma\langle m_{\nu} \rangle^2  =
\frac{\sum{w_i \langle m_{\nu} \rangle_i^2}}{\sum{w_i}} \pm (\sum
w_i)^{-1/2},
 \label{eq:6}
\end{equation}

\noindent where

\begin{equation}
w_i=1/\sigma\langle m_{\nu} \rangle_i^2.
 \label{eq:7}
\end{equation}

In this approach negative values of the square values of the neutrino mass $\langle m_{\nu} \rangle^2_i$ can appear due to possible negative values of the peaks area. Finally, a combined limit on the neutrino mass can be calculated as the square root of the $\lim \langle m_{\nu} \rangle^2$, obtained from the values of $\langle m_{\nu} \rangle^2$ and $\sigma\langle m_{\nu} \rangle^2$.

This method reduces the impact of experiments with high background on
the combined neutrino mass limit. This approach can be applied to
extract a combined limit even from experiments with rather
different characteristics (e.g., bolometers, HPGe detectors,
Xe-filled scintillators, etc).

Below we will show how the discussed methods can be applied to the case of CUPID-like experiments.

\section{Results and discussion}
 \label{sec:3}

\subsection{Input data}
 \label{sec:31}

The various contributions to the background in a large scintillating and Cherenkov-radiation bolometric experiment and in the CUORE experiment are discussed respectively in Ref. \cite{Artusa:2014} and Ref. \cite{CUORE-BM}. According to the discussions in these works, we will assume that the external contributions to the background (external $\gamma$ radiation, muons, neutrons) can be reduced to below 10$^{-4}$ counts/yr/keV/kg by proper material selection and shielding. A simililarly low contribution will be provided by the bulk crystal radioactivity. The dominant source of background will be then related to contamination of close structures, typically the surface of the crystals and of the holders. Presently, $\alpha$ particles generated at the surface of the crystal holders are indeed the dominant background source in CUORE, at the level of 10$^{-2}$ counts/yr/keV/kg \cite{CUORE,CUORE-BM}. Switching on particle identification capability by the detection of light emitted by the crystals, the residual level of background is estimated by the simulations based on the CUORE background model and reported in Ref. \cite{Pavan-BG}. The remaining background events are determined by $\beta$'s and $\gamma$'s emitted by close contamination of the crystals and details of the set-up. The most challenging contaminant nuclei are the $\beta$ active $^{208}$Tl ($Q_{\beta}=4999$ keV, descendant of $^{228}$Th in the $^{232}$Th family) and $^{214}$Bi ($Q_{\beta}=3269$ keV, descendant  of  $^{226}$Ra in the $^{238}$U family). Both $Q_{\beta}$ values exceed the $Q_{2\beta}$ of the CUPID\=/4 isotopes of interest. 

The characteristics of a possible CUPID\=/1 set-up with $^{82}$Se, $^{100}$Mo, $^{116}$Cd or $^{130}$Te are summarized in Table \ref{tab:1}. The masses of the detectors and the numbers of nuclei should be divided by 4 to obtain the corresponding values for CUPID\=/4. We assume 10 yr of data taking in both configurations. We have not tried to optimize the sharing among the different isotopes, as this work intends just to show in general the potential of a multi-isotope bolometric approach. It is likely that, after the conclusion of the current CUPID R\&D phase, we will find out that the optimum crystal size and shape as well as the technical performances of the various detectors depend on the compound type: therefore, a sharing based on an equal number of crystals could not be the best choice. CUPID-2 or CUPID-3 options could also result more convenient. The choice of the isotopes can also depend on the future results of CUORE: a larger statistics will lead to a better definition of the background model and this can influence the $0\nu2\beta$ candidate selection, given the different values of the $Q_{2\beta}$'s.

\begin{table*}[t]
\caption{Characteristics of a possible CUPID set-up with $^{82}$Se,
$^{100}$Mo, $^{116}$Cd and $^{130}$Te in the CUPID\=/1 option
assuming 1300 detector modules with $\oslash5\times5$ cm crystals enriched to 100\% in the isotope of interest. The considered parameters are the FWHM energy resolution at $Q_{2\beta}$ (FWHM), the detection efficiency ($\varepsilon$), the detector mass ($M$) and the number of nuclei of interest ($N$).}
\footnotesize 
\begin{center}
\begin{tabular}{|c|c|c|c|c|c|c|}
 \hline
 Candidate       & $Q_{2\beta}$ (keV)    &Crystal             &FWHM            & $\varepsilon$  &$M$   &$N$  \\
 ~              & \cite{Wang:2017}      & compound                     & (keV) & ~    &(kg) & ~ \\
 ~              & ~                     & ~                     & ~  & ~             & ~      & ~ \\
 ~              & ~                     & ~                     & ~ & ~             &  ~        & ~ \\
 \hline
  $^{82}$Se     & 2997.9(5)             & Zn$^{82}$Se           & 13.8              & 0.76          & 672       & $2.75\times 10^{27}$  \\
 \hline
  $^{100}$Mo    & 3034.36(17)           & Li$_2$$^{100}$MoO$_4$   & 5.4               & 0.80          & 389       & $1.32\times 10^{27}$  \\
 \hline
  $^{116}$Cd    & 2813.49(13)           & $^{116}$CdWO$_4$      & 5.5               & 0.90          & 1008      & $1.67\times 10^{27}$  \\
 \hline
  $^{130}$Te    & 2527.51(1)            & $^{130}$TeO$_2$       & 5.0               & 0.87          & 771       & $2.87\times 10^{27}$  \\
 \hline

\end{tabular}
\end{center}
\label{tab:1}
\end{table*}

\subsection{Sensitivity of CUPID\=/1 experiments}
\label{sec:32}

\begin{figure}[b]
\centering
\includegraphics[width=0.48\textwidth]{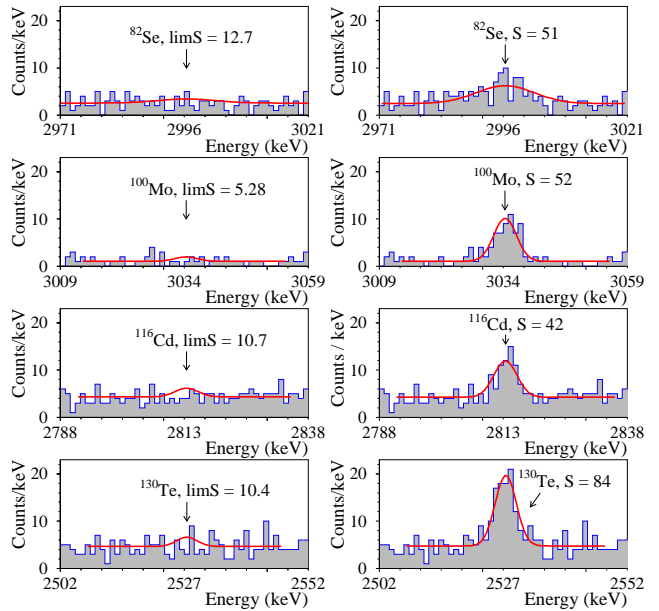}
\caption{Generated energy spectra for Zn$^{82}$Se,
Li$_2$$^{100}$MoO$_4$, $^{116}$CdWO$_4$ and $^{130}$TeO$_2$
detectors for the CUPID\=/1 option, assuming the background simulated in Ref.
\cite{Pavan-BG} (see also text and Table \ref{tab:1} for details).
The figures at the left column show the ``no effect observed" case, while those
at the right column illustrate the effects for a neutrino mass
of 0.05 eV calculated with the matrix elements from Ref. \cite{Barea:2015}, the
phase space factors from Ref. \cite{Kotila:2012} and the value of the
axial vector coupling constant $g_A=1.27$. Values of the excluded
$\lim S$ and peaks areas $S$ are in counts.}
\label{fig:1}
\end{figure}

\begin{figure}[thb]
\centering
\includegraphics[width=0.48\textwidth]{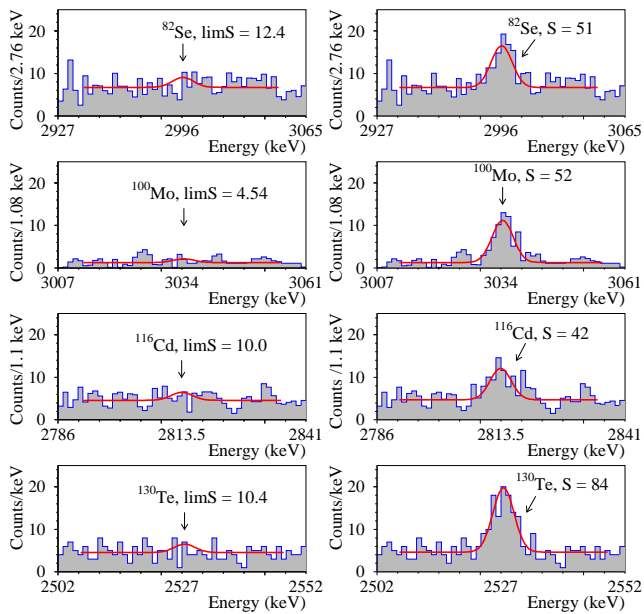}
\caption{The same spectra as in Fig. \ref{fig:1} are reported, but binning is units of energy resolution (1~bin~=~FWHM(keV)/5). This approach leads to a slightly better sensitivity, as discussed in the text.}
\label{fig:1a}
\end{figure}

First we estimate the sensitivity to the $0\nu2\beta$ decay of
$^{82}$Se, $^{100}$Mo, $^{116}$Cd and $^{130}$Te in the CUPID\=/1 option, assuming the simulated background reported in Ref. \cite{Pavan-BG}.\footnote{Strictly speaking, the background counting rate in Ref. \cite{Pavan-BG} was estimated for TeO$_2$ crystals. We use the simulation results also for the other compounds (taking into account of course the different $Q_{2\beta}$'s) assuming that the background level does not depend  significantly on the detector material, as expected if the contamination is mainly located in the holders.} Ten thousands of randomly generated energy
spectra were fitted by the maximum-likelihood method\footnote{We
have applied the likelihood fit taking into account that
approximation by the least squares method is not robust enough in
the case of low statistics data, since it ignores zero entry
bins.} to estimate the average errors of the peak area $\sigma(S_i)$ for
each nucleus. Examples of the fits are presented in Fig.
\ref{fig:1} (energy spectra with $S_i\approx0$ and $\sigma (S_i)$
equal to the average values for each detector were chosen for the
presentation). The values of $\lim S_i$ were estimated by using
the Feldman-Cousins procedure \cite{Feldman:1998} from the $\sigma(S_i)$ values
assuming $S_i=0$ counts. A slightly higher sensitivity was achieved by using energy spectra with bins in units of energy resolution as it is shown in Fig. \ref{fig:1a}.\footnote{In this Figure and in all the following ones where spectra are shown the binning is FWHM(keV)/5.} This improvement can be explained by an increase of statistics thanks to the binning in units of energy resolution. The results are presented in Table
\ref{tab:2}.\footnote{Similar estimations of $S_{\sigma}$ values can be obtained by the simple equation $S_{\sigma}\approx \sqrt{I_{BG}\times t \times M \times 1.75\times FWHM}$. However, this simple estimate becomes less relevant in case of lower statistics (even in the CUPID-1 option in case of shorter data-taking time).} The neutrino mass limits here and below have been
calculated by using the matrix elements from Ref. \cite{Barea:2015}, the
phase space factors from Ref. \cite{Kotila:2012} and the value of the
axial vector coupling constant $g_A=1.27$. We would like to stress
that the choice of the matrix elements from Ref. \cite{Barea:2015} does not
mean that we give preference to these calculations, but it is used only to illustrate our approach. In Section \ref{sec:35} we will give a picture of the CUPID\=/1 and CUPID\=/4 sensitivity to neutrino mass using other calculations too. In the right
column of Fig. \ref{fig:1} positive signals for the effective
neutrino mass $\langle m_{\nu} \rangle=0.05$ eV are shown (we have
used the same NME calculations to estimate the numbers of counts
as for the limit estimations above).

\begin{table*}[t]
\caption{Sensitivity of CUPID\=/1 experiments with
Zn$^{82}$Se, Li$_2$$^{100}$MoO$_4$, $^{116}$CdWO$_4$ and
$^{130}$TeO$_2$ detectors assuming the background in the ROI simulated in
Ref. \cite{Pavan-BG} (denoted as $I_{BG}$, counts/yr/keV/kg). The
limits on the number of events, the half-lives and the neutrino
mass are given at 90\% C.L. The neutrino mass limits are
calculated by using the matrix elements from Ref. \cite{Barea:2015},
the phase space factors from Ref. \cite{Kotila:2012} and the value of
the axial vector coupling constant $g_A=1.27$.}
\footnotesize
\begin{center}
\begin{tabular}{|l|l|l|l|l|l|}
 \hline
 Isotope        & $I_{BG}$              & $\sigma_S$    & $\lim S$  & $\lim T_{1/2}$    & $\lim \langle m_{\nu} \rangle$  \\
 ~              & ~                     & (counts)      & (counts)  & (yr)              & (eV) \\
 ~              &  ~                    & ~             & ~         & ~                 & ~ \\
 \hline
  $^{82}$Se     & $3.8\times 10^{-4}$   & 7.58          & 12.4      & $1.2\times 10^{27}$ & 0.025 \\
 \hline
  $^{100}$Mo    & $3.0\times 10^{-4}$   & 2.77          & 4.54      & $1.6\times 10^{27}$ & 0.015 \\
 \hline
  $^{116}$Cd    & $4.3\times 10^{-4}$   & 6.11          & 10.0      & $1.0\times 10^{27}$ & 0.025 \\
 \hline
  $^{130}$Te    & $6.0\times 10^{-4}$   & 6.36          & 10.4      & $1.7\times 10^{27}$ & 0.018 \\
 \hline

\end{tabular}
\end{center}
\label{tab:2}
\end{table*}

\subsection{Sensitivity of CUPID\=/4 and combined limits on the neutrino mass}
 \label{sec:33}

The sensitivity of CUPID\=/4 was estimated by using both the sum of counts in the ROI and the weighted averages methods.

Values of $S$ and $\sigma(S)$ can be estimated by using formula (\ref{eq:5}) and by a fit of the sum spectrum of several detectors. It should be noted that one can sum
spectra measured with different nuclei (and therefore with
different $Q_{2\beta}$ values). For this purpose the spectra can
be binned equally (e.g., 1 keV per channel, as in Fig.~\ref{fig:1}) or in units of energy resolution (as in Fig.~\ref{fig:1a}). They can then be shifted in
energy so that the decay energies of all the nuclei are in
the same channel. Binning in units of energy resolution improves statistics and looks a preferable approach in case of different energy resolutions. Below we will use energy spectra binned in units of energy resolution.

\begin{figure}[t]
\centering
\includegraphics[width=0.48\textwidth]{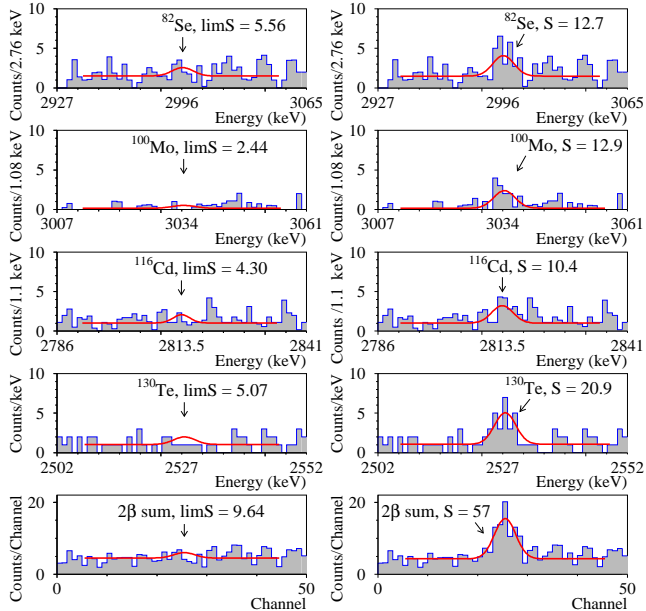}
 \caption{Generated energy spectra for Zn$^{82}$Se,
Li$_2$$^{100}$MoO$_4$, $^{116}$CdWO$_4$ and $^{130}$TeO$_2$
detectors for the CUPID\=/4 option and sum energy spectra,
assuming the background levels of Ref. \cite{Pavan-BG}. The figures at the left column
show the ``no effect observed" case, while those at the right column
illustrate effects for a neutrino mass of 0.05 eV calculated with
the matrix elements from Ref. \cite{Barea:2015}, the phase space factors
from Ref. \cite{Kotila:2012} and the value of the axial vector coupling
constant $g_A=1.27$. Values of excluded $\lim S$ and peaks areas
$S$ are in counts.}
 \label{fig:2}
 \end{figure}

The generated energy spectra of the four sets of detectors and the sum
spectrum are shown in Fig. \ref{fig:2}. The results of the fits are
presented in Fig. \ref{fig:2} and Table \ref{tab:3}. An estimation by
using formula (\ref{eq:5}) gives $\lim \langle m_{\nu}
\rangle=0.020$ eV. Then the sum spectrum was fitted by the sum of four
Gaussian functions with energy resolution FWHM=5 channels.
This approach gives almost the same result: $\lim \langle m_{\nu}
\rangle=0.021$ eV. The excluded peaks for the four detectors and
for the sum spectrum are shown in Fig. \ref{fig:2}.

The best limit $\lim \langle m_{\nu} \rangle=0.018$ eV was
obtained with the second method by using the values of weighted
errors of squares of neutrino mass $\sigma \langle m_{\nu}
\rangle^2$ listed in Table \ref{tab:3}.

Energy spectra generated for a Majorana neutrino mass of 0.05 eV
and their fits are given in the right column of Fig. \ref{fig:2}.
One can see that an effect for $\langle m_{\nu} \rangle=0.05$ eV will
be reliably observed both by CUPID\=/1 and CUPID\=/4 with the background levels extracted from Ref. \cite{Pavan-BG}. However, only the simultaneous observation in more than one isotope --- with excesses of counts in the correct energy positions --- would give overwhelming evidence for the occurrence of $0\nu2\beta$ decay, providing a cross check of the results in the moment itself of the discovery. 

\begin{table*}[t]
\caption{Sensitivity of the CUPID\=/4 experiment with
Zn$^{82}$Se, Li$_2$$^{100}$MoO$_4$, $^{116}$CdWO$_4$ and
$^{130}$TeO$_2$ detectors assuming the level of background
simulated in Ref. \cite{Pavan-BG} (denoted as $I_{BG}$,
counts/yr/keV/kg). The errors $\sigma\langle m_{\nu} \rangle^2$
and the neutrino mass limits were calculated by using the matrix elements from
Ref. \cite{Barea:2015}, the phase space factors from Ref. \cite{Kotila:2012}
and the value of the axial vector coupling constant $g_A=1.27$.}
\footnotesize
\begin{center}
\begin{tabular}{|l|l|l|l|l|l|l|l|}
 \hline
 Isotope        & Approach to estimate & $I_{BG}$              & $\sigma_S$    & $\lim S$  & $\lim T_{1/2}$    & $\lim \langle m_{\nu} \rangle$    & $\sigma\langle m_{\nu} \rangle^2$   \\
 ~              & $\langle m_{\nu} \rangle$ combined limit &~                     & (counts)      & (counts)  & (yr)              & (eV)                              & (eV$^2$) \\
 ~              &  ~ & ~                    & ~             & ~         & ~                 & ~                                 & ~ \\
 \hline
  $^{82}$Se     & ~ & $3.8\times 10^{-4}$   & 3.39          & 5.56      & $6.5\times 10^{26}$ & 0.033                           & 0.00067 \\
 \hline
  $^{100}$Mo    & ~ & $3.0\times 10^{-4}$   & 1.49          & 2.44      & $7.5\times 10^{26}$ & 0.022                           & 0.00029 \\
 \hline
  $^{116}$Cd    & ~ & $4.3\times 10^{-4}$   & 2.62          & 4.30      & $6.1\times 10^{26}$ & 0.032                           & 0.00063 \\
 \hline
  $^{130}$Te    & ~ & $6.0\times 10^{-4}$   & 3.09          & 5.07      & $8.5\times 10^{26}$ & 0.025                           & 0.00037\\
 \hline
 ~ & Estimation of peak & ~                     & 5.49          & 9.00      & ~                 & 0.020                             & ~ \\
 ~ & area by Eq. (\ref{eq:5}) (Sec.~\ref{sec:21}) &  ~             & ~             & ~         & ~                 & ~                                 & ~ \\
\hline
 ~ & Estimation of peak & ~                     & 5.88          & 9.64      & ~                 & 0.021                             & ~ \\
 ~ & area by fit (Sec.~\ref{sec:21})        &  ~                    & ~             & ~         & ~                 & ~                                 & ~ \\
 \hline
 ~ & Method described   & ~                     & ~             & ~         &                   & 0.018    $\Longleftarrow$         & 0.00020 \\
 ~ &      in Sec.~\ref{sec:22}    & ~                     & ~             & ~         &                   & ~                                 & ~ \\
 \hline

\end{tabular}
\end{center}
\label{tab:3}
\end{table*}

\subsection{Sensitivity of ``zero-background" experiments}
 \label{sec:34}

\begin{figure}[h]
\centering
\includegraphics[width=0.48\textwidth]{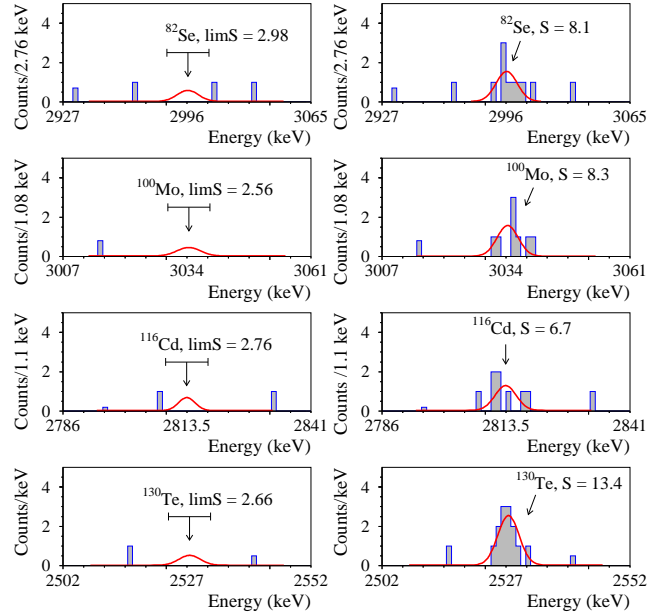}
 \caption{Generated energy spectra for Zn$^{82}$Se,
Li$_2$$^{100}$MoO$_4$, $^{116}$CdWO$_4$ and $^{130}$TeO$_2$
detectors for the CUPID\=/1 option, assuming a level of
background of $4\times 10^{-6}$ counts/yr/keV/kg. The figures at the left column
show the ``no effect observed" case. The energy intervals where
the numbers of background counts $N_{BG}$ were estimated (the
$N_{BG}$ values are given in Table \ref{tab:4}) are shown. The
figures at the right column illustrate effects for a neutrino mass
of 0.02 eV calculated with the matrix elements from Ref. \cite{Barea:2015}, the
phase space factors from Ref. \cite{Kotila:2012} and the value of the
axial vector coupling constant $g_A=1.27$. Values of excluded
$\lim S$ and peaks areas $S$ are in counts.}
 \label{fig:3}
 \end{figure}

\begin{figure}[h]
\centering
\includegraphics[width=0.48\textwidth]{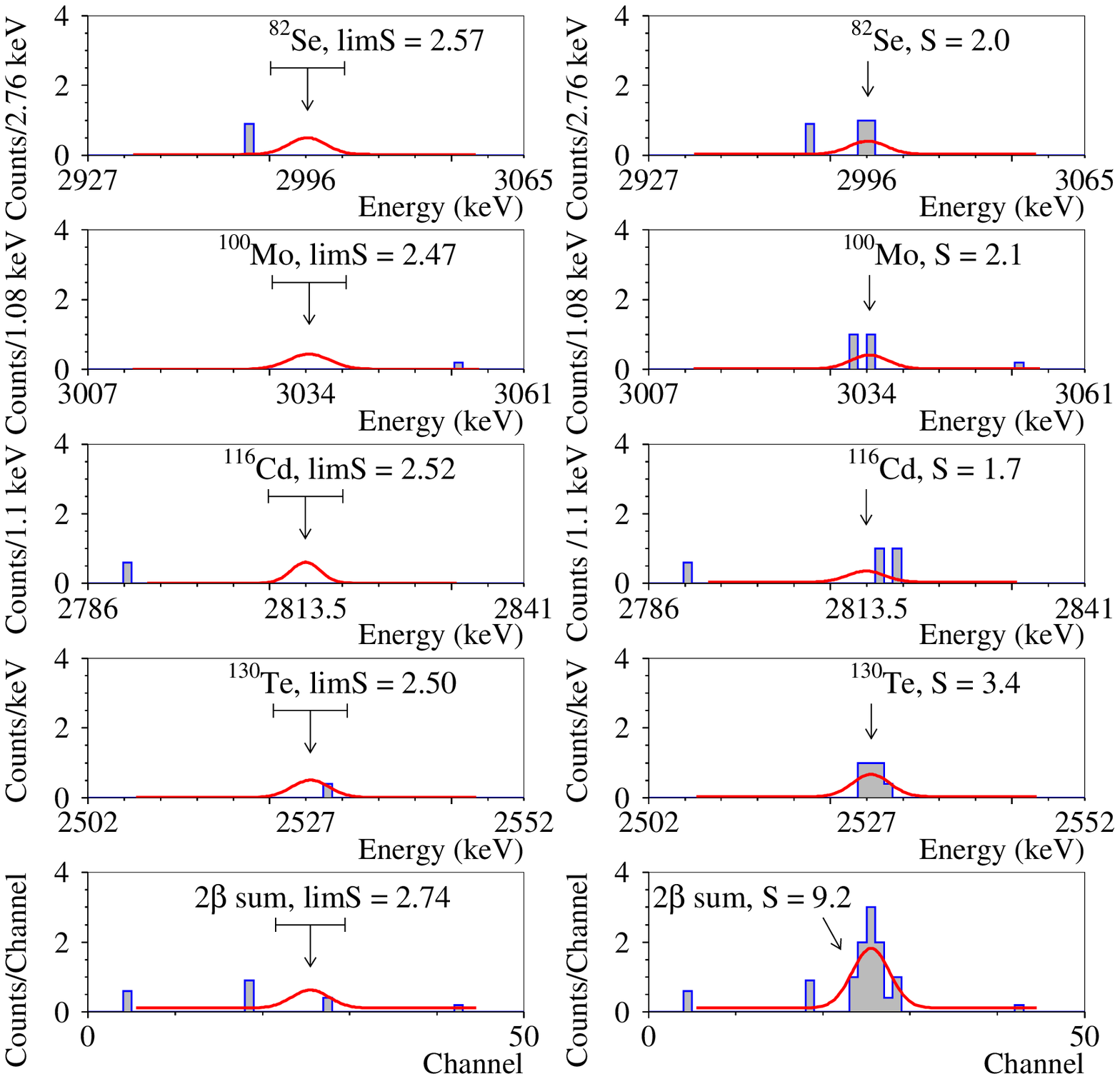}
 \caption{Generated energy spectra for Zn$^{82}$Se,
Li$_2$$^{100}$MoO$_4$, $^{116}$CdWO$_4$ and $^{130}$TeO$_2$
detectors for the CUPID\=/4 option and sum energy spectra,
assuming a level of background of $4\times 10^{-6}$
counts/yr/keV/kg. The figures  at the left column show the ``no effect observed"
case. The energy intervals where the numbers of background counts
$N_{BG}$ were estimated (the $N_{BG}$ values are given in Table
\ref{tab:5}) are shown. The figures at the right column illustrate
effects for a neutrino mass of 0.02 eV calculated with the matrix
elements from Ref. \cite{Barea:2015}, the phase space factors from
Ref. \cite{Kotila:2012} and the value of the axial vector coupling
constant $g_A=1.27$. Values of excluded $\lim S$ and peaks areas
$S$ are in counts.}
 \label{fig:4}
 \end{figure}

We will now assume that the background will be suppressed down to the level of $4\times10^{-6}$ counts/yr/keV/kg. This value is very challenging and we use it here just to simulate the case of a ``zero background experiment" both in the CUPID\=/1 and CUPID\=/4 options. Generated energy spectra for Zn$^{82}$Se, Li$_2$$^{100}$MoO$_4$, \linebreak[4] $^{116}$CdWO$_4$ and
$^{130}$TeO$_2$ detectors for the CUPID\=/1 option are
presented in Fig. \ref{fig:3}, while the energy spectra for the
CUPID\=/4 configuration are shown in Fig. \ref{fig:4}. A fit
of such statistically poor data is questionable. Thus, we have
estimated the sensitivity of this thought experiments in accordance
with the procedure reported in Ref. \cite{Feldman:1998} for an expected background and no
true signal (see Table XII in Ref. \cite{Feldman:1998}). The estimations of the experimental sensitivity for the CUPID\=/1 option are reported in Table \ref{tab:4}. The expected background $N_{BG}$ for each detector and for the sum spectrum was evaluated in
$\approx\pm 2\sigma$ energy intervals (the intervals are shown in Fig. \ref{fig:3}
and \ref{fig:4}). The estimations of the CUPID\=/4
experimental sensitivity are presented in Table \ref{tab:5}.

\begin{table}[htb]
\caption{Sensitivity of CUPID\=/1 experiments with
Zn$^{82}$Se, Li$_2$$^{100}$MoO$_4$, $^{116}$CdWO$_4$ and
$^{130}$TeO$_2$ detectors assuming a level of background of
$4\times 10^{-6}$ counts/year/kg/keV. The expected background in the energy intervals $\approx\pm 2\sigma$ are denoted as
$N_{BG}$. The neutrino mass limits $\lim \langle m_{\nu} \rangle$
were calculated by using the matrix elements from
Ref. \cite{Barea:2015}, the phase space factors from Ref. \cite{Kotila:2012}
and the value of the axial vector coupling constant $g_A=1.27$.}
\footnotesize
\begin{center}
\begin{tabular}{|l|l|l|l|l|}
 \hline
 Isotope        & $N_{BG}$  & $\lim S$  & $\lim T_{1/2}$        & $\lim \langle m_{\nu} \rangle$ \\
 ~              & (counts)  & (counts)  & (yr)                  & (eV)                           \\
 ~              &  ~        & ~         & ~                     & ~                              \\
 \hline
  $^{82}$Se     & 0.63      & 2.98      & $4.6\times 10^{27}$   & 0.012                          \\
 \hline
  $^{100}$Mo    & 0.14      & 2.56      & $2.7\times 10^{27}$   & 0.011                          \\
 \hline
  $^{116}$Cd    & 0.38      & 2.76      & $3.6\times 10^{27}$   & 0.013                          \\
 \hline
  $^{130}$Te    & 0.26      & 2.66      & $6.2\times 10^{27}$   & 0.009                          \\
 \hline

\end{tabular}
\end{center}
\label{tab:4}
\end{table}

\begin{table}[h]
\caption{Sensitivity of the CUPID\=/4 experiment with
Zn$^{82}$Se, Li$_2$$^{100}$MoO$_4$, $^{116}$CdWO$_4$ and
$^{130}$TeO$_2$ detectors assuming a level of background of
$4\times 10^{-6}$ counts/year/kg/keV. The numbers of background events in the energy
intervals $\approx\pm 2\sigma$ for each detector and for the sum spectrum (last line) are denoted as $N_{BG}$.
The neutrino mass limits $\lim \langle m_{\nu} \rangle$ were
calculated by assuming no true signal (see discussion in Secs.~\ref{sec:21} and ~\ref{sec:31}) and assuming the matrix elements from Ref. \cite{Barea:2015},
the phase space factors from Ref. \cite{Kotila:2012} and the value of
the axial vector coupling constant $g_A=1.27$.}
\footnotesize
\begin{center}
\begin{tabular}{|l|l|l|l|l|}
 \hline
 Isotope        & $N_{BG}$ & $\lim S$  & $\lim T_{1/2}$        & $\lim \langle m_{\nu} \rangle$ \\
 ~              & (counts) & (counts)  & (yr)                  & (eV)                           \\
 ~              & ~   & ~         & ~                     & ~                              \\
 \hline
  $^{82}$Se     & 0.16      & 2.57      & $1.3\times 10^{27}$   & 0.023                          \\
 \hline
  $^{100}$Mo   & 0.04      & 2.47      & $7.1\times 10^{26}$   & 0.022                          \\
 \hline
  $^{116}$Cd   & 0.09      & 2.52      & $9.9\times 10^{26}$   & 0.025                          \\
 \hline
  $^{130}$Te    & 0.07      & 2.50      & $1.6\times 10^{27}$   & 0.018                          \\
 \hline
Combined  & 0.36     & 2.74  & ~ & 0.011                          \\
$ \langle m_{\nu} \rangle$ limit & ~      & ~      & ~    & ~                          \\
 \hline

\end{tabular}
\end{center}
\label{tab:5}
\end{table}

The energy spectra for the Majorana neutrino mass $\langle
m_{\nu} \rangle=0.02$ eV are presented in the right columns of
Figs. \ref{fig:3} and \ref{fig:4}. The effect can be detected in
both CUPID configurations.

\subsection{Dependence on nuclear matrix elements calculations}
 \label{sec:35}

\begin{figure}[!h]
\centering
\includegraphics[width=0.4\textwidth]{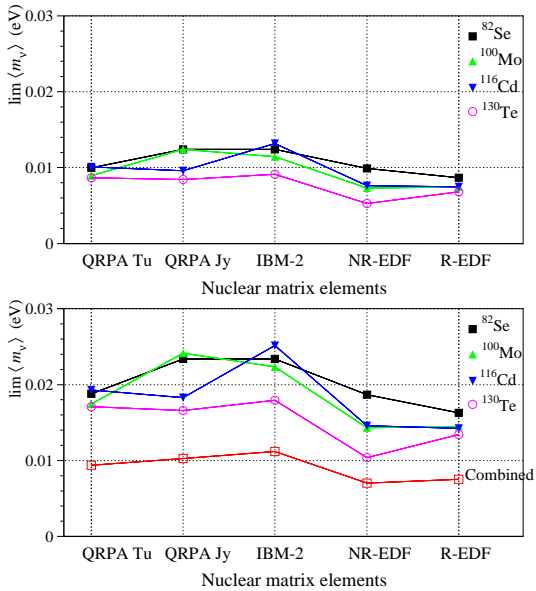}
 \caption{Sensitivity to the neutrino mass of $0\nu2\beta$ experiments with Zn$^{82}$Se,
Li$_2$$^{100}$MoO$_4$, $^{116}$CdWO$_4$ and $^{130}$TeO$_2$
detectors in the CUPID\=/1 (upper figure) and CUPID\=/4
(lower figure) options, assuming a level of background of $4\times
10^{-6}$ counts/yr/keV/kg. The values of $\lim \langle
m_{\nu}\rangle$ were estimated by using the phase space factors
from Ref. \cite{Kotila:2012}, the value of the axial vector coupling
constant $g_A=1.27$, and the NME calculated in the framework of
different nuclear models: QRPA Tu \cite{Simkovic:2013}; QRPA Jy \cite{Hyvarinen:2015}; IBM-2 \cite{Barea:2015}; NR-EDF \cite{Vaquero:2013}; and R-EDF \cite{Yao:2015}.}
 \label{fig:5}
 \end{figure}

The dependence of the CUPID thought experiment sensitivity to the
effective Majorana neutrino mass on the NME calculations is shown
in Fig. \ref{fig:5}. The experimental conditions are the same as those
reported in Tables \ref{tab:4} and \ref{tab:5}. The values of
$\lim \langle m_{\nu}\rangle$ were estimated by using the phase
space factors from Ref. \cite{Kotila:2012}, the axial vector coupling
constant $g_A=1.27$, and the NME discussed in a recent review \cite{Engel:2017} (however, we have considered only the works where all the ``CUPID nuclei'' ---~$^{82}$Se, $^{100}$Mo, $^{116}$Cd and $^{130}$Te~--- were calculated): the different versions of the quasiparticle random-phase approximation from the T\"ubingen (QRPA Tu) \cite{Simkovic:2013} and Jyv\"askyl\"a (QRPA Jy) \cite{Hyvarinen:2015}; the microscopic interacting boson model IBM-2 \cite{Barea:2015}; and energy density functional theory (EDF), non-relativistic (NR-EDF) \cite{Vaquero:2013} and relativistic (R-EDF) \cite{Yao:2015}. 

\section{Conclusions}
\label{sec:concl}

Two approaches to estimate the Majorana neutrino \linebreak[4] mass from
$0\nu2\beta$ decay experiments with several nuclei are proposed.
The former method uses a sum of counts in the region of interest, while
the latter one estimates the Majorana neutrino mass from weighted
averages and errors of neutrino mass squares. The former approach is
especially effective in case of similar detector performance
and can be used even in the ``zero background" conditions with a
very poor statistics, while the latter approach allows one to estimate
the neutrino mass also from experiments with completely different
detectors. 

By applying these methods, we have shown that the sensitivity to the neutrino mass that can be achieved
in a single bolometric experiment with several nuclei (e.g., with
$^{82}$Se, $^{100}$Mo, $^{116}$Cd, and $^{130}$Te) is similar to
the sensitivity of a technically equivalent experiment based on only one of these nuclei and contained in the same cryostat volume. Actually, Tables \ref{tab:2} and \ref{tab:4}, as well as Fig.~\ref{fig:5}, show that the pure $^{130}$Te option gives better results. We have to consider however that predictions of background are less safe for this nucleus, since this is the only one for which the signal is below the end point of the bulk of the natural $\gamma$ radioactivity (2615 keV). A possible way to mitigate the effects of a residual background component from external $\gamma$s is to place the TeO$_2$ crystals in the central part of the array so that the other crystals act as an active shield for the $^{130}$Te section. 

It should be stressed that a similar data analysis can be applied not only to the Majorana neutrino mass assessment in the framework of the mass mechanism involving light neutrino exchange, but also to the estimation of physics parameters related to other possible extensions of the SM that predict $0\nu2\beta$ decay.

Since a multi-isotope bolometric experiment is not less sensitive than a single-isotope one to the physics parameters that rule $0\nu2\beta$ decay (once that background and NME uncertainties are taken into account), we believe that it represents by far a better option, for a number of fundamental and practical reasons we have extensively discussed in the previous sections. This multi-isotope approach can be naturally implemented in the framework of CUPID, which can be based in principle on very similar detectors containing different candidates and exploiting well-established technologies. We emphasize that a competitive single experiment combining results from several candidates with similar sensitivities represents a new original idea in the $0\nu2\beta$ decay search strategy, which currently can be effectively implemented only in the bolometric approach.

Finally, it should be noted that a multi-isotope experiment could allow for precise measurements of the $2\nu2\beta$ decay half-lives, which are important nuclear data also for the development of a model aiming at accurate calculations of the neutrinoless mode of the transition.

\section{Acknowledgements}

The authors thank Dr. V.V.~Kobychev for the Monte-Carlo simulations of
$0\nu2\beta$ decay detection efficiencies of Li$_2$MoO$_4$ and
CdWO$_4$ detectors. The researches were supported in part by the
joint scientific project ``Development of Cd-based scintillating
bolometers to search for neutrinoless double-beta decay of
$^{116}$Cd'' in the fra\-me\-work of the PICS (Program of
International Cooperation in Science) of CNRS in years 2016--2018.
F.A.D. gratefully acknowledges support from the Scientific
High-Level Scholarship Commission of the French Embassy in Kyiv
(Ukraine) and the ``Jean d'Alembert'' Grants program of the
University of Paris-Saclay. F.A.D. and V.I.T. were supported in
part by the IDEATE International Associated Laboratory (LIA) and by the project ``Investigation of neutrino and weak interaction in double beta decay of $^{100}$Mo'' in the framework of the Programme ``Dnipro'' based on Ukraine-France Agreement on Cultural, Scientific and  Technological Cooperation.


\end{document}